\documentclass[%
 reprint,
groupedaddress,
showpacs,
 amsmath,amssymb,
 aps,
 prd,
 longbibliography,
 lengthcheck,%
]{revtex4-1}

\usepackage{graphicx}
\usepackage{dcolumn}
\usepackage{bm}
\usepackage{hyperref}
\usepackage{float}

\begin{document}


\title{Highly relativistic spinning particle \\ starting near
$r_{ph}^{(-)}$ in a Kerr field}


\author{Roman Plyatsko, Oleksandr Stefanyshyn, and Mykola Fenyk}
\affiliation{Pidstryhach Institute for Applied Problems in
Mechanics and Mathematics\\ Ukrainian National Academy of Sciences, 3-b Naukova Str.,\\
Lviv, 79060, Ukraine}


\date{\today}

\begin{abstract}
Using the Mathisson-Papapetrou-Dixon (MPD) equations, we
investigate the trajectories of a spinning particle starting near
$r_{ph}^{(-)}$ in a Kerr field and moving with the velocity close
to the velocity of light ($r_{ph}^{(-)}$ is the Boyer-Lindquist
radial coordinate of the counter-rotation circular photon orbits).
First, as a partial case of these trajectories, we consider the
equatorial circular orbit with $r=r_{ph}^{(-)}$. This orbit is
described by the solution that is common for the rigorous MPD
equations and their linear spin approximation. Then different
cases of the nonequatorial motions are computed and illustrated by
the typical figures. All these orbits exhibit the effects of the
significant gravitational repulsion that are caused by the
spin-gravity interaction. Possible applications in astrophysics
are discussed.
\end{abstract}


\maketitle

\section{ Introduction}

The geodesics in a Kerr metric are considered in the classical
books on general relativity [1--3]. Some recent papers are devoted
to more detailed study of geodesics on Kerr's black hole with the
aim to elucidate the mechanism of jet formation [4], and to
analyze the possibility of particle acceleration to arbitrary high
energy [5]. The complete sets of analytic solutions of the
geodesic equation in axially symmetric space-time are given in
[6]. However, the description of particle motion by geodesics is
restricted to a spinless particle. The motion of a spinning test
particle is described by the Mathisson-Papapetrou-Dixon equations
[7--9]:
\begin{equation}\label{1}
\frac D {ds} \left(mu^\lambda + u_\mu\frac {DS^{\lambda\mu}}
{ds}\right)= -\frac {1} {2} u^\pi S^{\rho\sigma}
R^{\lambda}_{\pi\rho\sigma},
\end{equation}
\begin{equation}\label{2}
\frac {DS^{\mu\nu}} {ds} + u^\mu u_\sigma \frac {DS^{\nu\sigma}}
{ds} - u^\nu u_\sigma \frac {DS^{\mu\sigma}} {ds} = 0,
\end{equation}
where $u^\lambda\equiv dx^\lambda/ds$ is the particle's
4-velocity, $S^{\mu\nu}$ is the tensor of spin, $m$ and $D/ds$
are, respectively, the mass and the covariant derivative with
respect to the particle's proper time $s$;
$R^{\lambda}_{\pi\rho\sigma}$ is the Riemann curvature tensor
(units $c=G=1$ are used). It is necessary to add a supplementary
condition to Eqs. (1), (2) in order to choose an appropriate
trajectory of the particle's center of mass. Most often the
conditions
\begin{equation}\label{3}
S^{\lambda\nu} u_\nu = 0
\end{equation}
and
\begin{equation}\label{4}
S^{\lambda\nu} P_\nu = 0
\end{equation}
are used, where
\begin{equation}\label{5}
P^\nu = mu^\nu + u_\lambda\frac {DS^{\nu\lambda}}{ds}
\end{equation}
is the 4-momentum. In practice, the condition for a spinning test
particle
\begin{equation}\label{6}
\frac{|S_0|}{mr}\equiv\varepsilon\ll 1
\end{equation}
must be taken into account [10], where  $|S_0|=const$ is the
absolute value of spin, $r$ is the coordinate distance of the
particle from the massive body.

After [7--9], Eqs. (1), (2) were obtained in many papers by
different approaches [11]. Also, this subject is of importance in
some recent publications [12].

In general, the solutions of Eqs. (1), (2) under conditions (3)
and (4) are different. However, in the post-Newtonian
approximation these solutions coincide with high accuracy [13],
just as in the case if the spin effects can be described by a
convergent in spin series, as some corrections to the
corresponding geodesic expressions [14]. Therefore, instead of
rigorous MPD Eqs. (1) their linear spin approximation
\begin{equation}\label{7}
m\frac D {ds} u^\lambda = -\frac {1} {2} u^\pi S^{\rho\sigma}
R^{\lambda}_{\pi\rho\sigma}
\end{equation}
is often considered. In this approximation condition (4) coincides
with (3).

 The effect of spin on the particle's motion in Kerr's field
has been studied since the 1970s [10,15,16]. In the past 10--12
years this subject gives rise to renewed interest [17--22],
particularly in the context of investigations of the possible
chaotic motions [17,19]. Also, these references provide a good
introduction concerning the MPD equations.

The purpose of this paper is to investigate more carefully the
world lines and trajectories of a spinning particle moving
relative to a Kerr source with the velocity close to the velocity
of light. We focus on the circular and close to circular orbits,
because just on these orbits one can expect the significant
effects of the spin-gravity interaction [10,23--25]. Indeed, these
orbits are of interest in the context of investigations of the
nongeodesic synchrotron electromagnetic radiation of highly
relativistic protons and electrons near black holes. Besides, it
is known that the highly relativistic circular orbits of a
spinless particle are of importance in the classification of all
possible geodesic orbits in a Kerr spacetime. Naturally, the
circular highly relativistic orbits of a spinning particle are of
importance in the classification of all possible significantly
nongeodesic orbits in this spacetime as well. Also, these orbits
are exclusive in the sense that they are described by the strict
analytical solutions of the MPD equations. The main features of
the spin-gravity interaction that are revealed on the circular and
close to circular orbits will be a good reference to further
investigations of most general motions of a spinning particle in a
Kerr spacetime.

We stress that MPD equations are the classical limit of the
general relativistic Dirac equation [26], new results in this
context are presented in [27]. Therefore, highly relativistic
solutions of the MPD equations stimulate the corresponding
investigations of the fermion's interaction with the strong
gravitational field.

The paper is organized as follows. Sec. 2 deals with the
relationships following from the MPD equations for the highly
relativistic equatorial circular orbits of a spinning particle in
Kerr's field, in the Boyer-Lindquist coordinates. The linear spin
MPD equations for any motions of a spinning particle are
considered in Sec. 3. The results of computer integration of these
equations for some significantly nongeodesic motions are presented
in Sec. 4. We conclude in Sec. 5.

Following [3], in this paper $r_{ph}^{(-)}$ notes the radial
coordinate of the photon circular orbits in the case of the
counter-rotation.

\section{Highly relativistic equatorial circular
orbits of spinning particles in a Kerr field according to
approximate and rigorous MPD equations}

In practical calculations it is convenient to represent
 the MPD equations through the spin 3-vector $S_i$, instead of the 4-tensor
$S^{\mu\nu}$, where by definition
  \begin{equation}\label{8}
S_i=\frac {1}{2} \sqrt{-g}\varepsilon_{ikl}S^{kl},
\end{equation}
where $g$ is the determinant of $g_{\mu\nu}$ , $\varepsilon_{ikl}$
is the Levi-Civita symbol (here and in the following, latin
indices run 1, 2, 3, and greek indices 1, 2, 3, 4, unless
otherwise specified). Then Eqs. (2) can be written as  [23]

\[u_{4} \dot S_i - \dot u_{4} S_i + 2(\dot u_{[4} u_{i]} - u^\pi
u_\rho \Gamma^\rho_{\pi[4} u_{i]})S_k u^k
\]
\begin{equation}\label{9}
 + 2S_n \Gamma^n _{\pi [4} u_{i]} u^\pi =0,
\end{equation}
where a dot denotes differentiation with respect to the proper
time $s$, and square brackets denote antisymmetrization of
indices; $\Gamma^\alpha_{\beta\gamma}$ are the Christoffel
symbols. Eq. (7) in terms of $S_i$ is
\[
m(\dot u^{\lambda} + \Gamma^{\lambda}_{\alpha\beta} u^\alpha
u^\beta)
\]
\begin{equation}\label{10}
+\frac {u^\pi}{    2u_4 \sqrt {-g}}(u_4 R^{\lambda}_{\pi ik} +
2u_i R^{\lambda}_{\pi k 4})\varepsilon^{ikl}S_l =0.
\end{equation}

In many papers the 4-vector of spin $s_\lambda$ is considered,
where
\[
s_{\lambda}=\frac{1}{2}\sqrt{-g}\varepsilon_{\lambda\mu\nu\sigma}u^{\mu}S^{\nu\sigma}.
\]
The following relationship holds:
\[
S_i=u_i s_4-u_4 s_i.
\]

Let us consider Eqs. (9), (10) for the Kerr metric using the
Boyer-Lindquist coordinates  $x^1=r, \quad x^2=\theta, \quad
x^3=\varphi, \quad x^4=t.$ Then the nonzero components of
$g_{\mu\nu}$ are
\[
g_{11}=-\frac{\rho^2}{\Delta}, \quad g_{22}=-\rho^2,
\]
\[
g_{33}=-\left(r^2+a^2+\frac{2Mra^2}{\rho^2}
\sin^2\theta\right)\sin^2\theta,
\]
\begin{equation}\label{11}
 g_{34}=\frac{2Mra}{\rho^2}\sin^2\theta, \quad
g_{44}=1-\frac{2Mr}{\rho^2},
\end{equation}
where
\[
 \rho^2=r^2+a^2\cos^2\theta, \quad \Delta=r^2-2Mr+a^2, \quad
0\le\theta\le\pi.
\]
[In the following we shall put $a\ge 0$, without any loss in
generality; the metric signature is -- -- -- +.] It is easy to
check that three equations from (9) have a partial solution with
$\theta=\pi/2 $, $S_1\equiv S_r=0$, $S_3\equiv S_\varphi=0$ and
the relationship for the nonzero component of the spin 3-vector
$S_2\equiv S_\theta=0$ is
\begin{equation}\label{12}
S_2=ru_4S_0,
\end{equation}
where $S_0$ is the constant of integration. The physical meaning
of this constant is the same as in the general integral of the MPD
equations [15]
\begin{equation}\label{13}
 S^2_0 = \frac 12 S_{\mu\nu}S^{\mu\nu}.
\end{equation}
We stress that relationship (12) is valid for any equatorial
motions ($\theta=\pi/2 $), with the spin orthogonal to the motion
plane ($S_i u^i=0$).

The possible equatorial orbits of a spinning particle are
described by Eq. (10). First, we shall consider the case of the
circular orbits with
 \begin{equation}\label{14}
  u^1=0 , \quad u^2=0, \quad u^3=const\neq0, \quad u^4=const\neq0.
\end{equation}
Investigating the conditions of existence of the equatorial
circular orbits for a spinning particle in Kerr's field we use
Eqs. (10), (12), and $ u^\mu u_\mu = 1 $. It is known that from
the geodesic equations in this field, the algebraic relationship
that follows determines the dependence of the velocity of a
spinless particle on the radial coordinate $r$ of the equatorial
circular orbit. Similarly, from the first equation of set (10)
using Eq. (14) we obtain the relationship for the equatorial
circular orbits of a spinning particle in Kerr's field as follows:
\[\Gamma^1_{33}(u^3)^2+ 2\Gamma^1_{34} u^3u^4+
 \Gamma^1_{44}(u^4)^2 - \frac{\Delta^2S_2}{r^4mu_4}
\]
\begin{equation}\label{15}
\times
[(u^3)^2R^4_{113}+u^3u^4(R^4_{114}-R^3_{113})-(u^4)^2R^3_{114}]=0.
\end{equation}
[By Eq. (14) other equations of set (10) are automatically
satisfied.] Taking into account Eq. (12) and the explicit
expressions for $\Gamma^{\lambda}_{\mu\nu}$ and
$R^{\lambda}_{\pi\rho\sigma}$ (see, e.g., [18]) from Eq. (15) we
obtain
  \[
  (Ma^2-r^3)(u^3)^2-2Mau^3u^4+M(u^4)^2-\frac{3MS_0}{mr^2}
  \]
  \begin{equation}\label{16}
 \times [a^2(r^2+a^2)(u^3)^2-(r^2+2a^2)u^3u^4+a(u^4)^2]=0.
   \end{equation}
The 4-velocity component $u^4$ can be expressed through $u^3$ from
the condition $ u^\mu u_\mu = 1 $ as follows:
\begin{equation}\label{17}
u^4=-\frac{g_{34}u^3}{g_{44}}+\sqrt{\frac{g^2_{34}-g_{33}g_{44}}{g^2_{44}}(u^3)^2+\frac{1}{g_{44}}}
\end{equation}
(just the sign "+" at the radical in Eq. (17) ensures the positive
value of $u^4$). Inserting expression (17) into Eq. (16) and
eliminating the radical by raising to the second power we get
 \[
 (u^3)^4(r^5[r(r-3M)^2-4Ma^2]+12\varepsilon Mar^4\Delta-9\varepsilon^2M^2r^4\Delta)
 \]
 \[+(u^3)^2(-2Mr^4(r-3M)+6\varepsilon Mar^3(r-3M)
 \]
\begin{equation}\label{18}
-9\varepsilon^2M^2r^3(r-2M)) + M^2(r-3\varepsilon a)^2=0,
\end{equation}
where, as in Eq. (6), $ \varepsilon=|S_0|/mr $. Without any loss
in generality we put $S_0>0$, then by Eq. (12) $S_2>0$.

So, the particle's angular velocity $\dot \varphi=u^3 $ on the
circular orbit with the radial coordinate $r$ must satisfy Eq.
(18).

Let us show that Eq. (18) provides known solutions. In the partial
case of a spinless particle ($\varepsilon=0$) from Eq. (18) we
have
 \begin{equation}\label{19}
 (u^3)^2=\frac{Mr(r-3M)+2Ma\sqrt{Mr}}{r^2[r(r-3M)^2-4Ma^2]}.
  \end{equation}
It follows from Eq. (19) that the velocity of such a particle on
the circular orbit is highly relativistic if the expression
 $r(r-3M)^2-4Ma^2$ is close to $0$. This fact is known from the
 analysis of the geodesic orbits in a Kerr field, as well as that
 just the roots of the equation
\[
r(r-3M)^2-4Ma^2=0
\]
determine the values of $r$ for the photon orbits.

If $ \varepsilon\neq 0 $ and the absolute value of the expression
$r^5[r(r-3M)^2-4Ma^2]$ in the factor of $(u^3)^4$ in Eq. (18) is
much greater than $12\varepsilon
Mar^4\Delta-9\varepsilon^2M^2r^4\Delta $, then it is easy to
verify that the corresponding roots of Eq. (18) describe the
circular orbits with the angular velocity which is close to the
angular velocity of the corresponding geodesic orbits due to $
\varepsilon\ll 1 $. More exactly, in this case we have
\begin{equation}\label{20}
|u^3|=\sqrt{\frac{Mr(r-3M)+2Ma\sqrt{Mr}}{r^2[r(r-3M)^2-4Ma^2]}}(1+O(\varepsilon)),
\end{equation}
where the main term is equal to the square root of the right-hand
side of Eq. (19).

Now we point out a case of interest with $\varepsilon\neq 0 $ that
is not described in the literature. Namely, it is not difficult to
calculate that for $ \varepsilon\neq 0 $ Eqs. (16), (18) have the
solutions which describe the highly relativistic circular orbits
with the values of $r$ that is equal or close to $r_{ph}^{(-)}$,
i.e., to the radial coordinate of the counter-rotation photon
circular orbits. For example, in the case of the maximum Kerr
field  ($a=M$) the orbits with $r=4M(1+\delta)$, where
$|\delta|\ll 1$, are highly relativistic, both for positive and
negative $\delta$. If $|\delta|\ll \varepsilon$, according to Eqs.
(16), (18) the values of $u^3$ on these orbits are determined by
the expression
\begin{equation}\label{21}
   u^3=-\frac1{3M\sqrt{6\varepsilon(1+4\delta/3\varepsilon)}}(1+O(\varepsilon)).
\end{equation}
[The choice of the sign in Eq. (21), $u^3<0$, is dictated by the
necessity to satisfy both Eq. (18) and (16); Eq. (18), as compared
to (16), has additional roots because of the operation of raising
to the second power.] Similarly as in the case of Eq. (20), it is
easy to check that if $\varepsilon\ll |\delta|$ it follows from
Eqs. (16), (18) the expression for $u^3$ which in the main term
coincides with the known analytic solution for the corresponding
geodesic circular orbit.

It follows from Eqs.
(16), (18) at $r=r_{ph}^{(-)}$ for any $0\le a\le M$ that
\begin{equation}\label{22}
u^3=-\frac{2(M/r_{ph}^{(-)})^{3/4}}{\sqrt{3\varepsilon}(r_{ph}^{(-)}-M)}(1+O(\varepsilon)).
\end{equation}
 It is known from the geodesic equations that the
 values of $r_{ph}^{(-)}$ increase monotonically from $3M$ at $a=0$ to  $4M$ at $a=M$.

Thus, according to Eq. (22) the expression for the angular
velocity $\dot \varphi=u^3 $ in the main term is proportional to
$1/\sqrt{\varepsilon}$, whereas the angular velocity in Eq. (20)
at $r=r_{ph}^{(-)}(1+\delta)$, $0<\delta\ll 1$ is proportional to
$1/\sqrt{\delta}$. Further details appear below in this Sec.

Using Eq. (22) we can estimate the value of the Lorentz
$\gamma$-factor, corresponding to the 4-velocity component $u^3$ ,
for different $a$. More exactly, we shall calculate the Lorentz
$\gamma$-factor from the point of view of an observer which is at
rest relative to a Kerr mass. According to the general expression
for the 3-velocity components $v^i$ we write [1]
 \begin{equation}\label{23}
v^i=\frac{dx^i}{\sqrt{g_{44}}}\left(dt+\frac{g_{4i}}{g_{44}}dx^i\right)^{-1}
\end{equation}
and for the second power of the velocity absolute value $|v|^2$ we
have
 \begin{equation}\label{24}
\quad |v|^2=v_iv^i=\gamma_{ij}v^iv^j,
\end{equation}
where $\gamma_{ij}$ is the 3-space metric tensor. The relationship
between $\gamma_{ij}$ and $g_{\mu\nu}$ is as follows:
\begin{equation}\label{25}
\gamma_{ij}=g_{ij}+\frac{g_{4i}g_{4j}}{g_{44}}.
\end{equation}
For the circular motions we have
 $ v^1=0, \quad v^2=0$, and according to Eq. (23)
\begin{equation}\label{26}
\quad
v^3=\frac{dx^3}{\sqrt{g_{44}}}\left(dt+\frac{g_{4i}}{g_{44}}dx^i\right)^{-1}=
\frac{u^3}{\sqrt{g_{44}}}\left(u^4+\frac{g_{43}}{g_{44}}u^3\right)^{-1}.
\end{equation}
By Eqs. (24)-(26) and the condition $ u^\mu u_\mu = 1 $, for the
$\gamma$-factor we write
\begin{equation}\label{27}
\gamma=\quad\frac1{\sqrt{1-v^2}}=\sqrt{
(u^3)^2\left(-g_{33}+\frac{g^2_{43}}{g_{44}}\right)+1}.
\end{equation}
Inserting the value $u^3$ from Eq. (22) into Eq. (27) we find in
the corresponding spin approximation
\begin{equation}\label{28}
\gamma=\frac{1}{\sqrt{3\varepsilon}}\left(\frac{M}{r_{ph}^{(-)}}\right)^{-1/4}\left(1-\frac{2M}{r_{ph}^{(-)}}\right)^{-1/2}.
\end{equation}
It follows from Eq. (28) that  $\gamma^2>>1$, i.e., the under
consideration circular motions are highly relativistic. Fig. 1
shows the dependence $\gamma/\gamma_0$ on $r_{ph}^{(-)}/M$, where
$\gamma_0$ is the $\gamma$-factor for $r_{ph}^{(-)}=3M$, that is,
at $a=0$.

\begin{figure}[H]
\centering
\includegraphics[width=7cm]{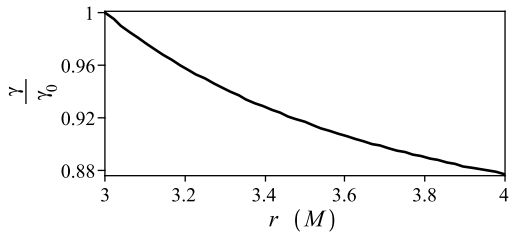}
\caption{\label{1} Ratio $\gamma$-factor at different $a$, from
$0<a\le M$, to $\gamma$-factor at $a=0$ vs. radial coordinate.}
\end{figure}

It is known that the important characteristic of the particle's
motion in the Kerr spacetime are its energy and angular momentum.
Let us estimate the conserved values of the energy $E$ and angular
momentum $J$ of a spinning particle on the above considered highly
relativistic circular orbits with Eqs. (21), (22).  The
expressions for these quantities are [16,28]
\begin{equation}\label{29}
E=mu_4+g_{34}u_{\mu}\frac {DS^{3\mu}} {ds}+g_{44}u_{\mu}\frac
{DS^{4\mu}} {ds}+\frac{1}{2}S^{\mu\nu}g_{\nu4,\mu},
\end{equation}
 \begin{equation}\label{30}
J=-mu_3-g_{33}u_{\mu}\frac {DS^{3\mu}} {ds}-g_{34}u_{\mu}\frac
{DS^{4\mu}} {ds}-\frac{1}{2}S^{\mu\nu}g_{\nu3,\mu}.
\end{equation}
By Eqs. (21), (29) for $r=4M(1+\delta)$, $|\delta|\ll \varepsilon$
we obtain
 \begin{equation}\label{31}
E_{spin}=\frac{m}{\sqrt{6\varepsilon}}.
\end{equation}
The energy of a spinless particle on the geodesic circular orbit
with  $r=4M(1+\delta)$, $0<\delta\ll 1 $ for $a=M$ is equal to
\begin{equation}\label{32}
E_{geod}=\frac{3m}{2\sqrt{10\delta}}.
\end{equation}
It follows from Eqs. (31), (32) that
\begin{equation}\label{33}
  E^2_{spin}/m^2 \gg 1 , \quad E^2_{geod}/m^2 \gg 1.
\end{equation}
At the same time according to (31), (32) for $0<\delta\ll
\varepsilon$ we have
\begin{equation}\label{34}
E^2_{spin}/E^2_{geod}=\frac{20\delta}{27\varepsilon}\ll 1.
\end{equation}
That is, the values of energy of the spinning and spinless
particles on the highly relativistic circular orbits with the same
$r=4M(1+\delta)$ in the maximal Kerr field can differ
significantly. It is easy to show that similar situation takes
place for all values $0\le a\le M$ with
$r=r_{ph}^{(-)}(1+\delta)$. In addition, one can estimate for
these circular orbits that according to Eq. (30)
$J^2_{spin}/J^2_{geod} \ll 1$.

As a result, the relationships $E^2_{spin}/E^2_{geod}\ll 1$ and
$J^2_{spin}/J^2_{geod} \ll 1$ following from Eqs. (29), (30) show
clearly that the corresponding solutions of the MPD equations
cannot be obtained in the framework of the analytic perturbation
approach to the dynamics of a classical spinning particle
developed in [20--22].

We stress that in this Sec. above we have considered the new
highly relativistic  circular solutions of the approximate MPD
Eqs. (2), (7). Let us show that these solutions satisfy the
rigorous MPD Eqs. (1), (2). Indeed, it follows from Eqs. (1) that
their terms, which were neglected in Eqs. (7), in the case of the
circular equatorial motions are presented in the first Eq. of set
(1) only. In metric (11) these terms can be written as
 \begin{equation}\label{35}
  c_1(u^3)^4+c_2(u^4)^4+c_3(u^3)^3u^4+c_4(u^3)^2(u^3)^2+c_5u^3(u^4)^3,
 \end{equation}
where
\[
c_1=S_0{\Delta}\frac{Ma}{r^7}(3r^2+a^2)(r^3-Ma^2),
\]
\[c_2=-S_0{\Delta}\frac{M^2a}{r^7}\left(1-\frac{2M}{r}\right),
\]
\[ c_3=S_0\frac{\Delta}{r^7}[r^5(r-3M)(r^3-3Ma^2)+4M^2a^4],
\]
\[c_4=S_0{\Delta}\frac{Ma}{r^7}\left(3r^3-11Mr^2-6Ma^2+\frac{2M^2a^2}{r}\right),
\]
\begin{equation}\label{36}
 c_5=-S_0\frac{M\Delta}{r^7}\left(r^3-3Mr^2-4Ma^2+\frac{4M^2a^2}{r}\right).
\end{equation}
First, we point out that according to Eq. (36) in the case of a
Schwarzschild's field for the circular orbit with $r=3M$ the all
coefficients $c_i$ are equal to 0. Therefore, in this case
expression (35) is equal to 0 identically, independently on the
explicit expressions for $u^3$, $u^4$. It means that the highly
relativistic circular orbit of a spinning particle with $r=3M$ in
a Schwarzschild's field is a common strict solution both of the
approximate MPD equations (2), (7) and the rigorous MPD equations
(1), (2). Second, it is not difficult to check that applying Eqs.
(17), (22) to expression (35) at $a\ne 0$ yields $0$ in the main
terms, i.e., within the accuracy of order $\varepsilon$.

Thus, in this Sec. we dealt with new partial solutions of the MPD
equations in a Kerr spacetime with Eqs. (12), (14), (22). All
highly relativistic orbits of a spinning particle described by
these solutions are circular and located in the space region with
$r=r_{ph}^{(-)}(1+\delta)$, $|\delta|\ll 1$. To study highly
relativistic orbits beyond this region it is necessary to carry
out the corresponding computer calculations. It is an aim of Secs.
3 and 4.

\section{Equations (7), (9) for any motions in a Kerr field }

 Now the point of interesting is the noncircular highly
 relativistic motions of a spinning particle that starts near
 $r_{ph}^{(-)}$ in a Kerr field. In particular, we shall
 consider the effect of the 3-vector $S_i$ inclination to the
 equatorial plane $\theta=\pi/2 $ on the particle's trajectory. In
 this case Eqs. (9) cannot be integrated separately from Eqs. (7).
 For computer integration, it is necessary to write the explicit
 form of  Eqs. (7), (9) in metric (11). It is convenient to use
 the dimensionless quantities $y_i $, where by definition
\[ y_1=\frac{r}{M},\quad
y_2=\theta,\quad y_3=\varphi, \quad y_4=\frac{t}{M},
\]
\[
      y_5=u^1,\quad y_6=Mu^2,\quad y_7=Mu^3,\quad y_8=u^4,
\]
\begin{equation}\label{37}
    y_9=\frac{S_1}{mM},\quad y_{10}=\frac{S_2}{mM^2},\quad
    y_{11}=\frac{S_3}{mM^2},
\end{equation}
and
\begin{equation}\label{38}
x=\frac{s}{M}, \quad \varepsilon_0= \frac{|S_0|}{mM}
\end{equation}
(in contrast to $\varepsilon$ from (6), that depends on $r$, here
$\varepsilon_0$ is defined to be $const$). Then it follows from
Eqs. (7), (9) the set of 11 first-order differential equations
\[
\dot y_1 = y_5,\quad \dot y_2 = y_6,\quad \dot y_3 = y_7,\quad
\dot y_4 = y_8,
\]
\[
   \dot y_5 = A_1,\quad \dot y_6 = A_2,\quad
\dot y_7 = A_3,\quad \dot y_8 = A_4,
\]
\begin{equation}\label{38}
 \dot y_9= A_5,\quad \dot y_{10} = A_6,\quad \dot y_{11} = A_7,
\end{equation} where a dot denotes differentiation with respect
to $x$, and $A_j$ ($j=1,...,7$) are some functions which depend on
$y_i$. Because the expressions for $A_i$ in general case of any
$a$ are too lengthy, here we write $A_i$ for the much simpler case
$a=0$:
\[
A_1=\frac{y_5^2}{y_1^2}q +\left(y_1y_6^2+y_1y_7^2\sin^2y_2
-\frac{y_8^2}{y_1^2}\right)q^{-1}
\]
\[
 + \frac{3}{y_1^3\sin y_2}(y_7y_{10}\sin^2y_2-y_6y_{11}),
\]
\[
A_2=y_7^2\cos y_2\sin
y_2-\frac{2y_5y_6}{y_1}-\frac{3y_7y_9}{y_1^3}\sin y_2,
\]
\[
A_3=-\frac{2y_5y_7}{y_1}-2y_6y_7\cot y_2+\frac{3y_6y_9}{y_1^3\sin
y_2},
\]
\[
A_4=-\frac{2y_5y_8}{y_1^2}q-\frac{3y_5}{y_1^3y_8\sin y_2}q^2
\]
\[
\times(y_7y_{10}\sin^2y_2-y_6y_{11}),
\]
\[
A_5=\frac{2}{y_1^2}y_5y_9q
+\frac{y_6y_{10}}{y_1}+\frac{y_7y_{11}}{y_1}
\]
\[
-(y_5y_9+y_6y_{10}+y_7y_{11})[A_1q
\]
\[
 -\frac{y_5}{y_8}A_4q-3\frac{y_5^2}{y_1^2}q^2
\]
\[
-y_1y_6^2-
y_1y_7^2\sin^2y_2+\frac{y_8^2}{y_1^2}]+\frac{y_9}{y_8}A_4,
\]
\[
A_6=-y_1y_6y_9\left(1-\frac{3}{y_1}\right)+\frac{y_5y_{10}}{y_1}\left(\frac{2}{y_1}q+1\right)
\]
\[
+ y_7y_{11}\cot
y_2+\frac{y_{10}}{y_8}A_4-(y_5y_9+y_6y_{10}+y_7y_{11})
\]
\[
\times [y_1^2A_2-\frac{y_1^2y_6}{y_8}A_4+2y_5y_6(y_1- q)
\]
\[
- y_1^2y_7^2\cos y_2\sin y_2],
\]
\[A_7=\frac{y_5y_{11}}{y_1}\left(\frac{2}{y_1}q+1\right)
\]
\[
-y_1y_7y_9\left(1-\frac{3}{y_1}\right)\sin^2 y_2- y_6y_{11}\cot
y_2
\]
\[-y_7y_{10}\cos y_2\sin
y_2-(y_5y_9+y_6y_{10}+y_7y_{11})\left[A_3y_1^2\sin^2y_2 \right.
\]
\[
 -A_4\frac{y_1^2y_7}{y_8}\sin^2y_2+2y_5y_7(y_1-q)\sin^2y_2
\]
\begin{equation}
\label{40}\left.+2y_1^2y_6y_7\cos y_2\sin
y_2\right]+\frac{y_{11}}{y_8}A_4,
\end{equation}
where
\[
q\equiv \left(1-\frac{2}{y_1}\right)^{-1}.
\]
According to Eqs. (13), (38) the expression for $\varepsilon_0^2$
can be written as
\[
\varepsilon_0^2=\frac{y_9^2}{y_8^2}q(1+qy_5^2) +
\frac{y_{10}^2}{y_1^2y_8^2}q^2(1+y_1^2y_6^2)
\]
\[
+\frac{y_{11}^2q^2}{y_1^2y_8^2\sin^2y_2}(1+y_1^2y_7^2\sin^2y_2)
\]
\begin{equation}
\label{41}
+\frac{2q^2}{y_8^2}(y_5y_6y_9y_{10}+y_5y_7y_9y_{11}+y_6y_7y_{10}y_{11}).
\end{equation}
Eqs. (39)--(41) are aimed at computer integration.

\section{Numerical results}

We present here the results of computer integration of Eqs. (39)
with (40), (41). All plots below are restricted to the domains of
validity of the linear spin approximation. That is, in these
domains the neglected terms of the rigorous MPD equations are much
less than the linear spin terms. We monitor errors of computing
using the conserved quantities: the absolute value of spin, the
energy and angular momentum (see Eqs. (13), (29), (30)).

Figs. 2--9 correspond to the case of Schwarzschild's field. All
plots start with the same initial values of the coordinates
$x^1=r, \quad x^2=\theta, \quad x^3=\varphi, \quad x^4=t$, namely,
at $3M, \quad 90^{\circ}, \quad 0^{\circ},$  and $0$
correspondingly. We do not vary the initial values of the
4-velocity components $u^2$ and $u^3$ as well. More exactly, we
put $u^2=0$ and
\begin{equation}
\label{42} u^3=-\frac{1}{3M\sqrt{2}}\sqrt{-1+\frac{\sqrt{12 +
\varepsilon_0^2}}{\varepsilon_0}},
\end{equation}
where expression (42) is the solution of Eq. (18) at $a=0$,
rigorous in $\varepsilon_0$. That is, the initial values of $u^2$
and $u^3$ are the same as for the equatorial circular orbit with
$r=3M$ (it is easy to check that expression (42) coincides with
(22) in the corresponding spin approximation). However, we vary
the initial inclination angle of the spin 3-vector to the plane
$\theta=90^{\circ}$, without change of the absolute value of spin
(Figs. 2--6), and add the small perturbation by the radial
velocity (Figs. 7--9). Without any loss in generality we put
$S_1>0$. For comparison, we present the corresponding solutions of
the geodesic equations that start with the same initial values of
the coordinates and velocity as the solutions of the equations for
a spinning particle. In all cases as a typical value we put
$\varepsilon_0=10^{-4}$.

Figs. 2--9 let us compare the world lines and trajectories of the
spinning and spinless particles in  Schwarzschild's field. Figs.
2, 6, 7, 9 exhibit the significant repulsive effects of the
spin-gravity interaction on the spinning particle. Due to the
repulsive action the spinning particle falls on the horizon
surface during longer time as compared to the spinless particle
(Fig. 2). Moreover, according to Figs. 6, 9 the considerable space
separation takes place within a short time, i.e., within the time
of the spinless particle's fall on the horizon.

The point of interest is the phenomenon when a spinning particle
orbits below the equatorial plane for some revolutions, Figs. 3, 8
(we recall that according to the geodesic equations similar
situation is impossible for a spinless particle). It is easy to
calculate that for $S_1<0$ a spinning particle can orbit above the
equatorial plane.

We stress that all graphs in Figs. 2--9 are interrupted beyond the
domain of validity of the linear spin approximation. By Figs. 3,
4, within the time of this approximation validity, the period of
the $\theta$-oscillation coincides with the period of the
particle's revolution by $\varphi$. Whereas on this interval the
value of the spin component $S_1$ is $const$ (Fig. 5), just as the
components $S_2$ and $S_3$ (the corresponding graphs are not
presented here). We point out that this situation differs from the
corresponding case of the circular motions of a spinning particle
that are not highly relativistic. Indeed, then the nonzero radial
spin component is not $const$ but oscillates with the period of
the particle's revolution by $\varphi$. Besides, in the last case
the mean level of $\theta$ coincides with $90^{\circ}$, whereas in
Figs. 3, 8 the mean values of $\theta$ are above $90^{\circ}$.
According to Figs. 3, 8 the amplitude of the $\theta$-oscillation
increases with the inclination angle. However, $\theta -
90^{\circ}$ is small even for the inclination angle that is equal
to $90^{\circ}$.

In the context of Figs. 3, 8 we point out an interesting
analytical result following from Eqs. (39)--(41). Namely, it is
not difficult to check that these equations are satisfied if
\[ y_1=\frac{3}{1-\delta_1},\quad
y_2=\arccos \delta_2,\quad y_3=0, \quad y_4=0,
\]
\[y_5=0, \quad y_6=0, \quad
y_7=\pm
\frac{1}{3\sqrt{6\delta_1}}(1-\delta_1)^{3/2}(1-\delta_2^2)^{-1/2},
\]
\[
y_8=\left(1-\frac{2}{y_1}\right)^{-1/2}\sqrt{1+y_1^2y_7^2(1-\delta_2^2)},
\]
\[
y_9=\pm
\frac{3\delta_2}{\sqrt{6\delta_1}}(1-\delta_1)^{-3/2}(1-\delta_2^2)^{-1/2},
\]
\begin{equation}\label{43}
 y_{10}=-y_9
\frac{3\delta_1}{\delta_2}(1-\delta_1)^{-1}(1-\delta_2^2)^{1/2},
\quad y_{11}=0,
\end{equation}
where $\delta_1$ and $\delta_2$ are some small constant values
such that $0<\delta_1\ll 1$, $0<|\delta_2|\ll 1$. According to Eq.
(41) the relationship between $\delta_1$, $\delta_2$ and
$\varepsilon$ from Eq. (6) is as follows:
\begin{equation}\label{44}
\varepsilon^2=3\delta_1^2+\delta_2^2.
\end{equation}
By notation (37) we conclude that the partial solution of Eqs.
(39)--(40), that is presented in Eq. (43), describes the highly
relativistic nonequatorial circular motion with
$r=3M(1-\delta_1)^{-1}, \quad \cos \theta=\delta_2, \quad \dot
\varphi=const, \quad S_1=const\ne 0, \quad S_2=const\ne 0, \quad
S_3=0$. That is, Eq. (43) shows the possibility of the spinning
particle solution that orbit permanently above or below the
equatorial plane, however, with the small value
$|\theta-90^{\circ}|$ only. One can verify that due to the small
values $\delta_1$, $\delta_2$ solution (43) is an approximate
solution of the rigorous MPD equations.

Just as in the case of  Schwarzschild's field, the set of Eqs.
(39) with the corresponding expressions for $A_i$ can be
integrated numerically in Kerr's field. Fig. 10, as an analogy of
Fig. 2, shows the plots of $r(s)$ for some values of the
inclination angle at $a=M$. Fig. 11 shows the dependence of the
amplitude and period of the $\theta$-oscillation on the Kerr
parameter $a$.  The main features of Figs. 4--9 are peculiar to
the corresponding plots for Kerr's field, that are not presented
here for brevity.

According to Figs. 2, 6, 10 when the inclination angle is equal to
$90^{\circ}$, i.e., when spin lies in the equatorial plane, so
that spin-gravity coupling is equal to $0$, the corresponding
plots coincide with the geodesics. It is an evidence of the
correct transition from solutions of the MPD equations to
geodesics.

Fig. 12 illustrates some typical cases of the equatorial motions
of a particle with fixed initial values of its coordinates and
velocity but with different absolute value of spin (the
inclination angle is equal to $0^{\circ}$) in Kerr's field at
$a=M$. All curves start from $r=4M$ with zero radial velocity and
the tangential velocity which at $\varepsilon_0=10^{-4}$ is needed
for the circular motion with $r=4M$. This circular motion is shown
by the horizontal line, whereas other curves represent the
noncircular motions at $\varepsilon_0<10^{-4}$ with the same
particle's initial values of coordinates and velocity. According
to Fig. 12 the circular orbit of a spinning particle with $r=4M$
monotonically trend to the corresponding noncircular geodesic
orbit if $\varepsilon_0$ trend to 0, i.e., the limiting transition
$\varepsilon_0\to 0$ is correct.

\begin{figure}[H]
\centering
\includegraphics[width=7cm]{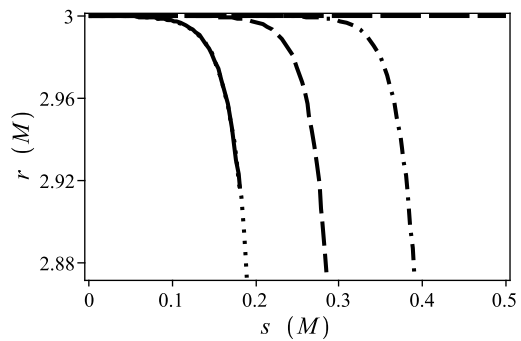}
\caption{\label{2} Radial coordinate vs. proper time for the
inclination angle $0^{\circ}$ (horizontal line $r=3M$),
$1^{\circ}$ (dash and dot line), $10^{\circ}$ (dash line), and
$90^{\circ}$ (solid line) at $a=0$, $\varepsilon_0=10^{-4}$. The
dot line corresponds to the geodesic motion with the same initial
values of the coordinates and velocity.}
\end{figure}

\begin{figure}[H]
\centering
\includegraphics[width=6.6cm]{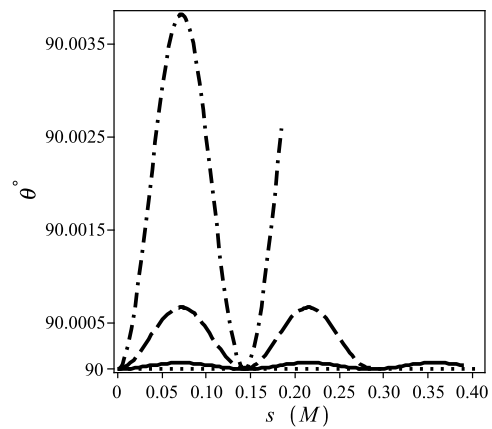}
\caption{\label{3} Graphs of the angle $\theta$ vs. proper time
for the inclination angle $0^{\circ}$ (horizontal line
$\theta=90^{\circ})$, $1^{\circ}$ (solid line), $10^{\circ}$ (dash
line), and $90^{\circ}$ (dash and dot line) at $a=0$,
$\varepsilon_0=10^{-4}$.}
\end{figure}

\begin{figure}[H]
\centering
\includegraphics[width=7.5cm]{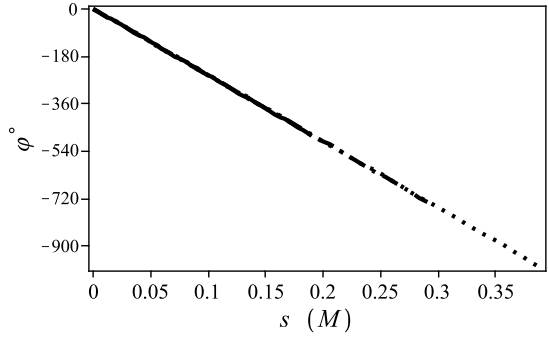}
\caption{\label{4} Graphs of the angle $\varphi$ vs. proper time
at $a=0$, $\varepsilon_0=10^{-4}$ for different values of the
inclination angle practically coincide with the corresponding
geodesic plot. The same feature takes place for the corresponding
graphs t vs. s that are not presented here for brevity.}
\end{figure}

\begin{figure}[H]
\centering
\includegraphics[width=7.5cm]{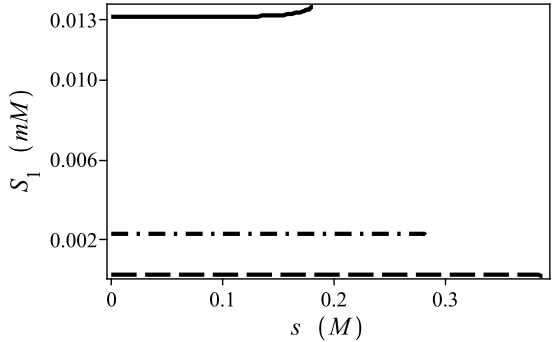}
\caption{\label{5} Graphs of $S_1$ vs. proper time for the
inclination angle  $1^{\circ}$ (dash line), $10^{\circ}$ (dash and
dot line), and $90^{\circ}$ (solid line) at $a=0$,
$\varepsilon_0=10^{-4}$.}
\end{figure}

\begin{figure}[H]
\centering
\includegraphics[width=6cm]{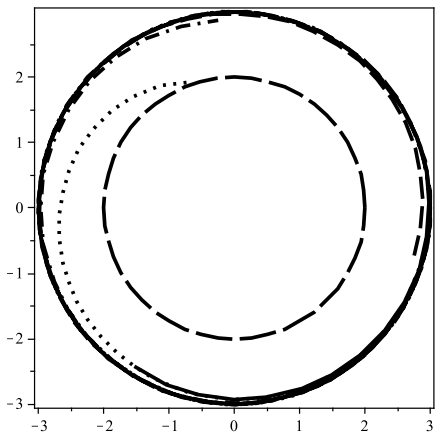}
\caption{\label{6} Trajectories of the spinning particle in the
polar coordinates for the inclination angle $0^{\circ}$ (circle
$r=3M$), $1^{\circ}$ (dash line), $10^{\circ}$ (dash and dot
line), and $90^{\circ}$ (solid line) at $a=0$,
$\varepsilon_0=10^{-4}$. The dot line corresponds to the geodesic
motion with the same initial values of the coordinates and
velocity. The circle $r=2M$ corresponds to the horizon line.}
\end{figure}

\begin{figure}[H]
\centering
\includegraphics[width=7cm]{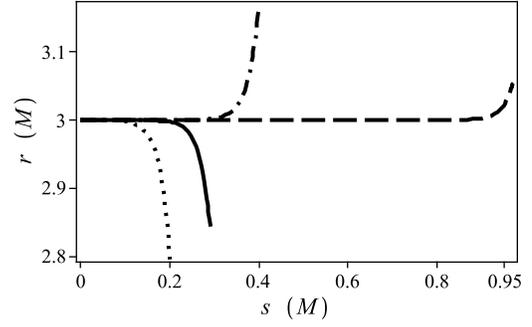}
\caption{\label{7} Radial coordinate vs. proper time for the
inclination angle $0^{\circ}$ (dash and dot line), $1^{\circ}$
(dash line), and $10^{\circ}$ (solid line) at $a=0$,
$\varepsilon_0=10^{-4}$, and the nonzero radial velocity
$dr/ds\approx 3.858\times 10^{-7}$. The dot line corresponds to
the geodesic motion with the same initial values of the
coordinates and velocity.}
\end{figure}

\begin{figure}[H]
\centering
\includegraphics[width=7cm]{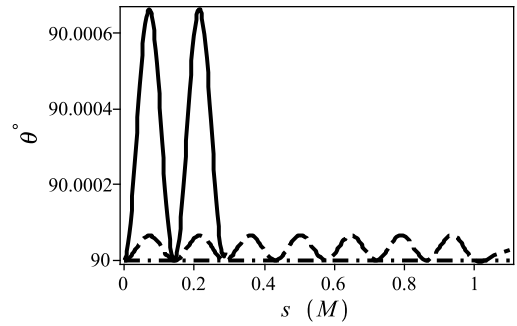}
\caption{\label{8} Graphs of the angle $\theta$ vs. proper time
for the inclination angle $0^{\circ}$ (horizontal line),
$1^{\circ}$ (dash line) and $10^{\circ}$ (solid line), at $a=0$,
$\varepsilon_0=10^{-4}$, and $dr/ds\approx 3.858\times 10^{-7}$.}
\end{figure}

\begin{figure}[H]
\centering
\includegraphics[width=6cm]{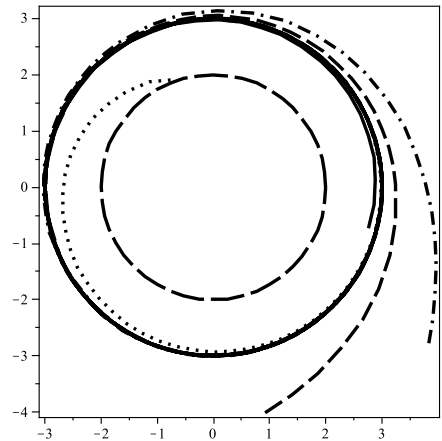}
\caption{\label{9} Trajectories of the spinning particle in the
polar coordinates for the inclination angle $0^{\circ}$ (dash and
dot line), $1^{\circ}$ (dash line), and $10^{\circ}$ (solid line)
at $a=0$, $\varepsilon_0=10^{-4}$, and $dr/ds\approx 3.858\times
10^{-7}$. The dot line corresponds to the geodesic motion with the
same initial values of the coordinates and velocity.}
\end{figure}

\begin{figure}[H]
\centering
\includegraphics[width=7cm]{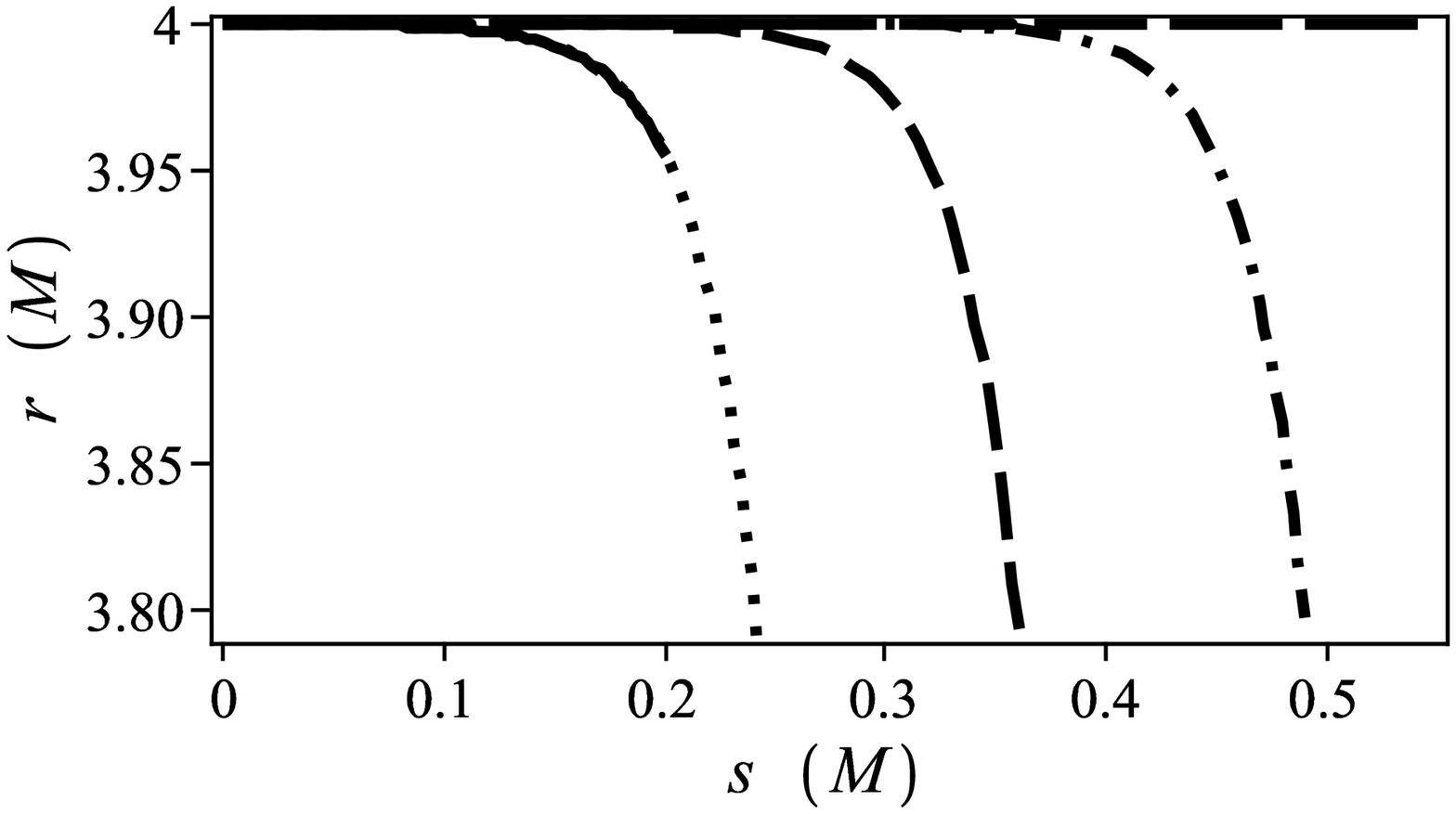}
\caption{\label{10} Radial coordinate vs. proper time for the
inclination angle $0^{\circ}$ (horizontal line $r=4M$),
$1^{\circ}$ (dash and dot line), $10^{\circ}$ (dash line), and
$90^{\circ}$ (solid line) at $a=M$, $\varepsilon_0=10^{-4}$. The
dot line corresponds to the geodesic motion with the same initial
values of the coordinates and velocity. }
\end{figure}

\begin{figure}[H]
\centering
\includegraphics[width=7cm]{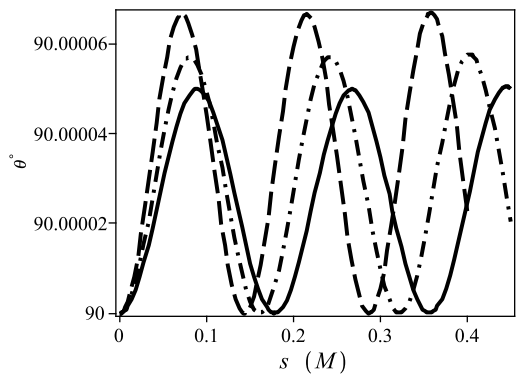}
\caption{\label{11} Graphs of the angle $\theta$ vs. proper time
for the inclination angle $1^{\circ}$ at $a=0$ (dash line),
$a=0.5M$ (dash and dot line), $a=M$ (solid line);
$\varepsilon_0=10^{-4}$.}
\end{figure}

\begin{figure}[H]
\centering
\includegraphics[width=6cm]{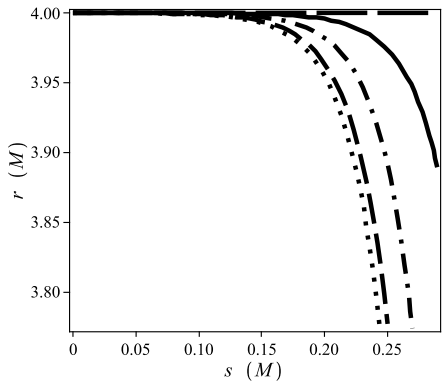}
\caption{\label{12} Radial coordinate vs. proper time for the
equatorial motions in Kerr's field at $a=M$ with different
absolute value of spin and the same (common) initial values of the
coordinates and velocity. The circular orbit with $r=4M$ at
$\varepsilon_0=10^{-4}$ is shown by the long dash line. The solid,
dash and dot, and dash lines describe the cases when
$\varepsilon_0$ is equal to $0.9\times 10^{-4}$, $0.6\times
10^{-4}$, and $0.2\times 10^{-4}$ correspondingly. The dot line
represents the geodesic motion with the same initial values of the
coordinates and velocity.}
\end{figure}

Finally, we remark on a simple conclusion following from the
nongeodesic curves of a spinning particle presented, in part, in
Figs. 2--12. Let us consider any point on the trajectory of a
spinning particle corresponding to its proper time $s_1>0$ (we
recall that all curves in Figs. 2--12 begin at $s=0$). Then the
geodesic curve can be calculated which starts just at this point
with the velocity that is equal to the velocity of the spinning
particle at the same point. Also, it is not difficult to estimate
the deviation of the pointed out nongeodesic curve from this
geodesic. In principle, it means that in our comparison of the
corresponding geodesic and nongeodesic curves we are not
restricted to the trajectories of a spinning particle which start
in the small neighborhood of the value $r=r_{ph}^{(-)}$ only
(naturally, here we are restricted to the domain of validity of
the linear spin approximation). This conclusion may be useful for
the generalization of the results obtained in this Sec. on other
motions of a spinning particle.

\section{Conclusions}

In this paper, using the linear spin approximation of the MPD
equations, we have studied the significantly nongeodesic highly
relativistic motions of a spinning particles starting near
$r=r_{ph}^{(-)}$ in Kerr's field. Some of these motions, namely
circular, are described by the analytical relationships following
directly from MPD equations in the Boyer-Lindquist coordinates.
Others, noncircular and nonequatorial, are calculated numerically.
For realization of these motions the spinning particle must
possess the obital velocity corresponding to the relativistic
Lorentz $\gamma$-factor of order $1/\sqrt{\varepsilon}$. All
considered cases of the spinning particle motion are within the
framework of validity of the test-particle approximation when
$\varepsilon\ll 1$.

The situation with a macroscopic test particle moving relative to
a massive body with $\gamma^2\gg 1$ is not realistic. However, the
highly relativistic values of the Lorentz $\gamma$-factor are
usual in astrophysics for the elementary particles. For example,
if $M$ is equal to three of the Sun's mass (as for a black hole),
then $\varepsilon_0$ for an electron is of order $0.4\times
10^{-16}$ and from Eq. (28) we have the $\gamma$-factor of order
$2\times 10^8$. Similarly, for a neutrino with the mass $\approx 1
eV$ we have $\varepsilon_0\approx 2\times 10^{-11}, \quad
\gamma\approx 3\times 10^5$.

We can expect the effects of the significant space separation of
some highly relativistic particles with different orientation of
spin. Indeed, the effects considered in this paper exhibit the
strong repulsive action of the spin-gravity interaction. For
another correlation of signs of the spin and the particle's
orbital velocity this interactions acts as an attractive force. In
general, the last case is beyond the validity of the linear spin
approximation. The nonlinear spin effects will be investigated in
another paper. Also, we plan to show that according to the MPD
equations significantly nongeodesic orbits of a highly
relativistic spinning particle, with the $\gamma$-factor of order
$1/\sqrt{\varepsilon}$, exist for the much wider space region of
the initial values of the particle's coordinates in a Kerr
spacetime than the orbits considered above. However, the
corresponding calculations are very complicated because this
result follows from the rigorous MPD equations (1), (2) only, and
is not common for the approximate equations (2), (7).

Naturally, it would be interesting to study the possible role of
the highly relativistic spin-gravity interaction in the jet
formation.


\begin{thebibliography}{99}

\bibitem{1} L. D. Landau and E. M. Lifshitz, {\it The classical theory of fields}
 (Addison-Wesley, Reading,
Massachusetts, 1971).
\bibitem{2} C. W. Misner, K. S. Thorne, and J. A. Wheeler,
 {\it Gravitation} (Freeman, San Francisko,
1973).
\bibitem{3} S. Chandrasekhar, {\it The Mathematical Theory of
Black Holes} (Oxford University Press, Oxford, 1983).
\bibitem{4} J. Gariel, M. A. H. MacCallum, G. Marcilhacy, and N. O.
Santos, Astron. and Astrophys. {\bf 515}, A15 (2010).
\bibitem{5} M. Banados, J. Silk, and S. M. West, Phys. Rev. Lett.
{\bf 103}, 111102 (2009); E. Berti, V. Cardoso, L. Gualtieri, F.
Pretorios, and U. Sperhake, Phys. Rev. Lett. {\bf 103}, 239001
(2009); T. Jacobson and T. Sotiriou, Phys. Rev. Lett. {\bf 104},
021101 (2010).
\bibitem{6} E. Hackmann, V. Kagramanova, J. Kunz, and C.
L{\"a}mmerzahl, Europhys. Lett. {\bf 88}, 30008 (2009).
\bibitem{7}
M. Mathisson, Acta Phys. Polon. {\bf 6}, 163 (1937).
\bibitem{8} A. Papapetrou, Proc. R. Soc. A {\bf 209}, 248 (1951).
\bibitem{9} W. G. Dixon, Proc. R. Soc. A {\bf 314}, 499 (1970);
Gen. Relativ. Gravitation {\bf 4}, 199 (1973); Philos. Trans. R.
Soc. A {\bf 277}, 59 (1974); Acta Phys. Pol. B. Proc. Suppl. {\bf
1}, 27 (2008).
\bibitem{10} R. Wald, Phys. Rev. D {\bf 6}, 406 (1972).
\bibitem{11} W. Tulczyjew, Acta Phys. Pol. {\bf 18}, 393 (1959);
A. Taub, J. Math. Phys. {\bf 5}, 112 (1964); P. Bartrum, Proc. R.
Soc. A {\bf 284}, 204 (1965); H. P. K{\"u}nzl, J. Math. Phys. {\bf
13}, 739 (1972); M. Omote, Progr. Theor. Phys. {\bf 49} 1559
(1973); S. Hojman, Phys. Rev. D {\bf 18}, 2741 (1978); A.
Balachandran, G. Marmo, B. Skagerstam, and A. Stern, Phys. Lett. B
{\bf 89} 199 (1980).
\bibitem{12} J. Natario, Commun. Math. Phys. {\bf 281}, 387
(2008); E. Barausse, E. Racine, and A. Buonanno, Phys. Rev. D {\bf
80}, 104025 (2009); J. Steinhoff and G. Sch{\"a}fer, Europhys.
Lett. {\bf 87}, 50004 (2009).
\bibitem{13} B. M. Barker and R. F. O'Connell, Gen. Rel. Grav. {\bf 4}, 193
(1973).
\bibitem{14}
A. N. Aleksandrov,  Kinem. Fiz. Nebesn. Tel. {\bf 7}, 13 (1991).
\bibitem{15} B. Mashhoon, J. Math. Phys. {\bf 12}, 1075 (1971).
\bibitem{16} K. P. Tod, F. de Felice, and M. Calvani, Nuovo Cim. B
{\bf 34}, 365 (1976).
\bibitem{17} S. Suzuki and K. Maeda, Phys. Rev. D {\bf 58}, 023005 (1998).
\bibitem{18} O. Semerak, Mon. Not. R. Astron. Soc. {\bf 308}, 863 (1999).
\bibitem{19} M. Hartl, Phys. Rev. D {\bf 67}, 024005 (2003); Phys. Rev. D {\bf 67}, 104023
(2003).
\bibitem{20} C. Chicone, B. Mashhoon, and B. Punsly, Phys. Lett. A
{\bf 343}, 1 (2005).
\bibitem{21} B. Mashhoon and D. Singh, Phys. Rev. D {\bf 74}, 124006
(2006).
\bibitem{22} D. Singh,
Phys. Rev. D {\bf 78}, 104028 (2008).
\bibitem{23}
R. Plyatsko, Phys. Rev. D. {\bf 58}, 084031 (1998).
\bibitem{24}
R. Plyatsko and O. Bilaniuk, Class. Quantum Grav. {\bf 18}, 5187
(2001).
\bibitem{25}
R. Plyatsko,  Class. Quantum Grav. {\bf 22}, 1545 (2005).
\bibitem{26}
S. Wong, Int. J. Theor. Phys.  {\bf 5}, 221 (1972); L. Kannenberg,
Ann. Phys. (N.Y.)  {\bf 103}, 64 (1977); R. Catenacci and M.
Martellini, Lett. Nuovo Cimento  {\bf 20}, 282 (1977); J.
Audretsch, J. Phys. A  {\bf 14}, 411 (1981); A. Gorbatsievich,
Acta Phys. Pol. B  {\bf 17}, 111 (1986); A. Barut and M. Pavsic,
Class. Quantum Grav.  {\bf 4}, 41 (1987).
\bibitem{27} F. Cianfrani and G. Montani, Europhys. Lett. {\bf
84}, 30008 (2008); Int. J. Mod. Phys. A {\bf 23}, 1274 (2008); Yu.
Obukhov, A. Silenko, and O. Teryaev, Phys. Rev. D {\bf 80}, 064044
(2009).
\bibitem{28}
R. Micoulaut, Z. Phys. {\bf 206}, 394 (1967).

\end{thebibliography}
\end{document}